\documentclass[paper]{JHEP3}
\usepackage{epsfig,graphicx,bm,amsmath}
\newcommand{\be}{\begin{equation}}
\newcommand{\ee}{\end{equation}}
\newcommand{\bea}{\begin{eqnarray}}
\newcommand{\eea}{\end{eqnarray}}
\newcommand{\beq}{\begin{equation}}
\newcommand{\eeq}{\end{equation}}
\newcommand{\beqa}{\begin{eqnarray}}
\newcommand{\eeqa}{\end{eqnarray}}
\newcommand{\beqar}{\begin{eqnarray*}}
\newcommand{\eeqar}{\end{eqnarray*}}
\newcommand{\beas}{\begin{eqnarray*}}
\newcommand{\eeas}{\end{eqnarray*}}
\usepackage{amsmath,amsfonts,amssymb,amsthm,amstext,amscd,eucal}
\newcommand{\ads}[1]{${\rm AdS}_{#1}$}

\preprint{IPM/P-2011/026}

\title{EVH Black Holes,  AdS$_3$ Throats and EVH/CFT Proposal}

\author{M.M.  Sheikh-Jabbari$^a$, Hossein Yavartanoo$^b$
\\
\\
$^a$ School of of Physics, Institute for Research in Fundamental Sciences (IPM),
P.O.Box 19395-5531, Tehran, Iran

\\
$^b$ Department of Physics, Kyung Hee University, Seoul 130-701, Korea}

\abstract{Within class of generic black holes there are extremal black holes (with vanishing Hawking temperature $T$) and vanishing horizon area $A_h$, but with finite $A_h/T$ ratio, the Extremal Vanishing Horizon (EVH) black holes. We  study the near horizon limit of a  four dimensional EVH black hole solution to a generic (gauged) Einstein-Maxwell dilaton theory and show that in the near horizon limit they develop a throat which is a pinching orbifold limit of  AdS$_3$. This is an extension of the well known result for extremal black holes the near horizon limit of which contains an AdS$_2$ throat. We show that in the near EVH near horizon limit the pinching AdS$_3$ factor turns to a pinching BTZ black hole and that
this near horizon limit is indeed a decoupling limit. We argue that the pinching AdS$_3$ or BTZ orbifold is resolved if the near horizon limit is accompanied by taking the 4d Newton constant $G_4$ to zero such that the Bekenstein-Hawking entropy $S=A_h/(4G_4)$ remains finite. We propose that in this limit the near horizon EVH black hole is dual to a 2d CFT. We provide pieces of evidence in support of the EVH/CFT correspondence and comment on its connection to the Kerr/CFT proposal and speculations how the EVH/CFT may be used to study generic e.g. Schwarzchild-type black holes.}


\begin{document}

%%%%%%%%%%%%%%%%%%%%%%%%%%%%%%%%%%%%%%%%%%%%%%%%%%%%
\section{Introduction}
%%%%%%%%%%%%%%%%%%%%%%%%%%%%%%%%%%%%%%%%%%%%%%%%%%%%

Einstein gravity coupled to various matter fields has black hole solutions. It is a well established fact that these black holes admit a thermodynamical description; one can associate a temperature $T$ with them which is the temperature of their Hawking radiation and an entropy $S$ which is the area of their horizon $A_h$  over $4G_N$, where $G_N$ is the Newton constant. Moreover,  black holes are specified by some quantum numbers, including their mass $M$, angular momenta $J_\alpha$ and electric or magnetic charges $Q_i$, and these quantum numbers, $T$ and $S$ satisfy the first law of thermodynamics. (In this description $T$ and $S$ are viewed as functions of these quantum numbers.) This thermodynamical description, although very interesting and insightful, once combined with Hawking's black hole evaporation analysis leads to a non-unitary description unless one can identify the black hole microstates and their underlying unitary statistical mechanical description.

Within the class of generic black holes there has been partial progress in identifying the microstates of supersymmetric black holes, which have  vanishing Hawking temperature \cite{{Strominger:1996sh},Reviews}.  This is usually done by first, realizing black holes as the gravity description of systems of branes and strings in  string theory setting and then use string theory dualities to map the system to a weakly coupled theory where the identification and counting of microstates can be performed.

Although a complete accounting of the quantum microstates of a generic black hole is not achieved yet, for which we presumably need a complete understanding of quantum theory of gravity, such an accounting has been obtained for many black hole solutions by identifying their underlying microstates with those of a dual two or higher dimensional CFT.  In most of the examples in which the identification has been done, black holes possesses an \ads{3}
throat in their near horizon limit and the degeneracy of their microstates can be captured by a
two-dimensional CFT using \ads{3}/CFT$_2$ duality.

 Besides  black holes with \ads{3} throat, there are some proposals towards the identification of microstates of extremal black holes. It has been shown that the near horizon geometry of extremal (not necessarily supersymmetric) black holes contain an \ads{2} throat \cite{Kunduri:2008rs,{Kunduri:2006ek}} and this fact, if we have a formulation of \ads{2}/CFT$_1$ duality (see \cite{Sen-AdS2/CFT1} for review on the progress in this direction), may be used for giving a statistical account of the black hole entropy.

It has been conjectured \cite{Kerr-CFT} that  an extremal Kerr black hole is dual to a chiral two dimensional CFT. This proposal is based on the observation that  the near horizon geometry of extremal Kerr black hole has an  $SL(2, {\mathbb R}) \times U(1)$ isometry group. Through
an asymptotic symmetry group analysis with certain boundary conditions, it is shown that the $U(1)$ part of this isometry enhances to a chiral Visaroso algebra with $c=12J$, $J$ being the angular momentum of the black hole. Finally, it is noted that the extremal black hole is at the Frolov-Thorne temperature $1/2\pi$. Then,
the entropy of the extremal black hole is equal to the entropy of the chiral 2d CFT, obtained using Cardy formula.
The conjecture has been extended to many other extremal black hole solutions \cite{Ext/CFT}.

Although the Kerr/CFT conjecture is very interesting, but, to be precise, it is rather a suggestion for a possible pair of theories dual to each other and  many things should  be understood to establish the proposal as a concrete duality. A precise identification of the proposed chiral CFT is still an open question and there has been arguments that the Extremal/CFT proposal does not have a dynamical content as the standard AdS/CFT and may only be used for reading the entropy (see \cite{DLCQ-CFT,Kerr-CFT-caveats} for a discussion on this point).

Here, we consider extremal ($T=0$) black holes which have also a vanishing horizon area ($A_h=0$) while  vanishing of $A_h$ and $T$ in the parameter space of black holes is  such that the $A_h/T$ ratio remains finite in the extremal limit. Black holes with this property will be called Extremal Vanishing Horizon (EVH) black holes. Explicitly,  suppose that we have a black hole specified with $n$ quantum numbers and charges. The EVH black hole parameter space, if exits, is then an $n-2$ dimensional hypersurface in this parameter space, associated with $A_h=0,\ T=0$. This hypersurface will be called the EVH hypersurface. Each point on the EVH hypersurface  corresponds to an EVH black hole.
The simplest  EVH black hole is the massless BTZ \cite{EVH-BTZ}, with $n=2$. The next simplest example with $n=3$ is 5d Kerr where one of the two angular momenta is vanishing. This example and its near horizon geometry was considered in \cite{bardeenhorowitz}.  Other examples of charged EVH black holes in AdS$_5$ and AdS$_4$ backgrounds, with respectively $n=4$ and $n=5$, where considered in \cite{Balasubramanian:2007bs,Fareghbal:2008ar,Fareghbal:2008eh}.

Within the class of EVH black holes there exists both supersymmetric and non-supersymmetric black holes, they can be static or stationary.
We should stress that the EVH black holes are \emph{not} small black holes (see \cite{Reviews} for discussion on small black holes), the condition of having fixed $A_h/T$ will discriminate the two. We will discuss this point further for the specific case of KK black holes in section 4.

One may  consider moving slightly away from the EVH hypersurface, then by definition, we get a \emph{near} EVH black hole with small $A_h$ and $T$ but with fixed $A_h/T$.  The near EVH black hole is hence specified by two parameters around a given EVH point.

In this work  we first set about refining arguments of \cite{Kunduri:2008rs} for the 4d EVH black hole  solutions to a generic (gauged) Einstein-Maxwell-dilaton (EMD) theory. In section 2, considering the most general 4d black hole with vanishing $T$ and $A_h$, and $A_h/T$ fixed, we find the most general form for the EVH black holes in 4d (gauged) EMD theory.

In section 3, we  prove that in general the near horizon limit of these EVH black holes has an \ads{3} throat. However, the circle in this \ads{3} does not have a full range of $[0,2\pi]$; it is a \emph{pinching \ads{3} orbifold} \cite{EVH-BTZ} (see also \cite{Terashima}). Moreover, we show that in the near horizon geometry of near EVH black holes the pinching \ads{3} throat turns into a pinching BTZ black hole. This pinching BTZ has parametrically the same entropy as the original near EVH black hole. EVH black holes are a special class of extremal black holes and one may  wonder if they show attractor behaviour \cite{attractor} and whether the entropy function method \cite{Sen:2005wa} works for them. This question, too, will be addressed in section 3.

In section 4, as specific examples  we consider the EVH KK black holes and review EVH black holes of 4d $U(1)^4$ gauged supergravity, discussed in \cite{Fareghbal:2008eh}. In section 5, we show that the near horizon limit discussed in section 3 is indeed a decoupling limit and propose that the pinching AdS$_3$ orbifold may be resolved once together with the near horizon limit we take $G_4\to 0$ limit such that $A_h/G_4$ remains finite. This paves the way to introduce the EVH/CFT correspondence: Physics on the near horizon (near) EVH black hole is described by a 2d CFT. We give a map between the parameters of the near horizon geometry and the 2d CFT quantum numbers. In section 6, we summarise our results and discuss extensions of the EVH/CFT correspondence and its connection to Kerr/CFT. In the appendix we have gathered some technical details of the calculations.

%%%%%%%%%%%%%%%%%%%%%%%%%%%%%%%%%%%%%%%%%%%%%%%%%%%%%%%%%%%%%%%%%%%%%%%%%%%%%%%%%%%%%%%%%%%%%%%%%%%%%%%%%%%5
%%%%%%%%%%%%%%%%%%%%%%%%%%%%%%%%%%%%%%%%%%%%%%%%%%%%%%%%%%%%%%%%%%%%%%%%%%%%%%%%%%%%%%%%%%%%%%%%%%%
\section{4d EVH black holes}
%%%%%%%%%%%%%%%%%%%%%%%%%%%%%%%%%%%%%%%%%%%%%%%%%%%%%%%%%%%%%%%%%%%%%%%%%%%%%%%%%%%%%%%%%%%%%%%%%%%%%
%%%%%%%%%%%%%%%%%%%%%%%%%%%%%%%%%%%%%%%%%%%%%%%%%%%%%%%%%%%%%%%%%%%%%%%%%%%%%%%%%%%%%%%%%%%%%%%%%%%%%%%%%%%5
Let us start with a generic four dimensional  Einstein-Maxwell-dilaton theory. We also assume scalar fields to have
a potential. This is the general structure of the bosonic part of 4d gauged supergravity theories. The action can be written as
\bea\label{Ein-Dil-Max}
S=-\frac{1}{\kappa_4^2}\int d^4x \sqrt{-g}\bigg[&& R-2G_{AB}\partial \Phi^A \partial \Phi^B-f_{IJ}(\Phi) F_{\mu\nu}^IF^{J\;\mu\nu} \cr\cr
&& -\frac{1}{2\sqrt{-g}}\epsilon_{\mu\nu\alpha\beta} \tilde{f}_{IJ}(\Phi)F^{I\;\mu\nu}F^{J\;\mu\nu} + V(\Phi) \bigg].
\eea
where $F^I_{\mu\nu}$ with $I=(1, \cdots N)$ are the gauge fields strengths, $\phi^A$ with
($A=1, \cdots, M$) are the scalar fields, and $k_4^2=16\pi G_4$; unless explicitly stated we will set $k_4=1$. The scalar fields $\Phi_A$, which in the absence of the potential $V(\Phi)$ are moduli,
determine the gauge coupling constants and $G_{AB}(\Phi)$ is the metric on the moduli
space. We use Gaussian units to avoid extraneous factors of $4\pi$ in the gauge fields. Varying the action we obtain the following equations of motion for the metric, scalar fields, and
the gauge fields:

\be
  R_{\mu\nu}-2G_{AB}\partial_{\mu}\Phi^A\partial_{\nu}\Phi^B + \frac{1}{2}g_{\mu\nu} V(\Phi)
  =f_{IJ}\left(2F^I_{\phantom{A}\mu\lambda}F^{J\phantom{\nu}\lambda}_{\phantom{J}\nu}-
    {\textstyle \frac{1}{2}}g_{\mu\nu}F^I_{\phantom{a}\alpha\lambda} F^{J \alpha\lambda} \right)
  \label{einstein}
\ee
\be
  \frac{1}{\sqrt{-g}}\partial_{\mu}(\sqrt{-g}G_{AB}\partial^{\mu}\Phi^B)
  =\frac{1}{4} \frac{\partial f_{IJ}}{\partial \Phi^A} F^I _{\phantom{I}\mu\nu} F^{J\, \mu\nu}
  +\frac{1}{8\sqrt{-g}}\frac{\partial \tilde f_{IJ}}{\partial \Phi^A}
  F^I_{\phantom{I}\mu\nu} F^J_{\phantom{J}\rho \sigma} \epsilon^{\mu\nu\rho\sigma} -\frac{dV}{d\Phi^A} \\
  \label{dilaton}
\ee
\be
  \partial_{\mu}\left[\sqrt{-g}\left(f_{IJ} F^{J\, \mu\nu}
      + \frac{1}{2\sqrt{-g}} {\tilde f}_{IJ}F^J_{\phantom{J}\rho\sigma}
      \epsilon^{\mu\nu\rho\sigma}\right) \right] =  0.
  \label{gaugefield}
\ee
The Bianchi identity for the gauge fields is
$F_{\phantom{I}\,[\mu\nu;\lambda]}^{I}=0$. For simplicity we restrict ourselves to the case with a single scalar and gauge field. It is straightforward to generalise our analysis  to the case with arbitrary number of  scalar and gauge fields. A stationary black hole solution (in ADM form) can be written as
\bea
\label{GenSol}
&& ds^2=-N^2(\rho,\theta)dt^2+ g_{\phi\phi}(\rho,\theta)\left(d\phi+N^{\phi}(\rho,\theta)dt\right)^2+g_{\rho\rho}(\rho,\theta)d\rho^2+g_{\theta\theta}(\rho,\theta) d\theta^2, \cr\cr
&& A=A_t(\rho,\theta)dt+A_{\rho}(\rho,\theta)d\rho+A_{\phi}(\rho,\theta) d\phi, \cr\cr
&& \Phi=\Phi(\rho,\theta).
\eea
The gauge field ansatz has been written in  $A_{\theta}=0$ gauge.

By eliminating the conical singularity in the Euclidean $(\tau = it, r)$ sector, we obtain the Hawking  temperature
\be
\label{temperatur}
T=\frac{1}{\Delta \tau} =\left(\frac{(N^2)'}{4\pi\sqrt{g_{\rho\rho}N^2}}\right)_{\rho=\rho_+}
\ee
where prime means derivative with respect to $\rho$ and $\rho_+$ is the location of the outer horizon. It is useful to rewrite the metric components in the form
\bea
\label{mu} N^2&=&(\rho-{\rho_+})(\rho-{\rho_-})\mu(\rho,\theta),  \\
\label{Np}N^{\phi}&=&-\omega+(\rho-\rho_+)\eta(\rho,\theta) ,     \\
\label{Lambda}g_{\rho\rho}&=& \frac{1}{(\rho-{\rho_+})(\rho-{\rho_-})\Lambda(\rho,\theta)}    \, ,
\eea
where we assume functions $\mu(\rho,\theta)$ and $ \Lambda(\rho,\theta)$ do not have zero in $(\rho_+,\infty)$. In addition,  having finite horizon angular velocity at the horizon, requires having finite  $\eta(\rho_+,\theta)$. The area of horizon $A_h$ can be expressed as
\be
\label{EVHen}
A_h=  {2\pi} \int_0^{\pi} \sqrt{g_{\theta\theta}^{(0)}(\theta)g_{\phi\phi}^{(0)}(\theta)} \;\; d\theta
\ee
where $g_{\theta\theta}^{(0)}$ and $g_{\phi\phi}^{(0)}$ are respectively values of $g_{\theta\theta}$ and $g_{\phi\phi}$ at horizon $\rho_+$. We also assume that these are analytic functions near the horizon, which means we can expand them as follows
\be
g_{\theta\theta}(\rho,\theta)= g_{\theta\theta}^{(0)}(\theta) + (\rho-\rho_+) g_{\theta\theta}^{(1)}(\theta) + \cdots , \;\;\;
g_{\phi\phi}(\rho,\theta)= g_{\phi\phi}^{(0)}(\theta) + (\rho-\rho_+) g_{\phi\phi}^{(1)}(\theta) + \cdots ,   \nonumber
\ee

The EVH point is defined by  $A_h\rightarrow 0$ and $T\rightarrow 0$ limit while  ${A_h}/{T}$ ratio is kept finite.
Let us first study implications of vanishing horizon area  $A_h\rightarrow 0$ limit, while  demanding the geometry to remain regular and non-trivial in this limit. This can be done if we scale $g_{\phi\phi}^{(0)} \rightarrow 0$. To be more precise we scale $g_{\phi\phi}^{(0)} \sim \tilde\epsilon^2$ for $\tilde\epsilon\rightarrow 0$, therefore $A_h\sim \tilde\epsilon$. It then suggests that the radius of outer horizon vanishes at this limit as well. Let us assume $\rho_+\sim \tilde\epsilon^s$, $s>0$. In addition we want to take extremal limit $T\rightarrow \tilde\epsilon$. This implies that the radius of inner horizon vanishes as well and therefore $\rho_+-\rho_-\sim \tilde\epsilon^v$ where $v\geq s$.

%The resulting geometry could be a naked singularity, but as we will see in more details for the example of KK black %hole in the next section, this could be a black hole solution where the horizon is pinched in two points. In other %words, except for two specific isolated points at fixed $\theta$ generically we do not have a naked singularity. The %EVH black hole solution to the theory \eqref{Ein-Dil-Max} away from these two values of $\theta$ is indeed a regular %geometry.

If we assume that after taking EVH limit $N^2$ is an analytic function of the radial coordinate near the horizon, which is located at $\rho=0$ we can expand it around $\rho=0$,
\be
\label{N2expan}
N^{2}(\rho,\theta)= \mu_0(\theta) + \rho \mu_1(\theta) + \rho^2 \mu_2(\theta) + \cdots\,.  \nonumber
\ee
Demanding  the geometry to be \emph{smooth} around $\rho=0$ (for generic values of $\theta$) implies that $N^2$ should vanish at the horizon (i.e. at $\rho=0$) and hence $\mu_0=0$.  If $\mu_1=0$ we could get a singularity at $\rho=0$ which is a naked singularity. However, since we desire to keep $\frac{A_h}{T}$ finite, from (\ref{temperatur}) it is easy to see that we need to keep $\mu_1\neq 0$.\footnote{$\mu_1(\theta)$ can still have some zeros at specific values of  $\theta$. At these specific values of $\theta$ we in fact have a naked singularity, while away from these values the geometry is smooth as we vary $\rho$. We can also keep $\Lambda (\rho,\theta)$ finite in this limit by a suitable choice of  the radial coordinate.}
It is hence more convenient to define
\be
N^2=\rho \tilde{\mu}(\rho,\theta)\,,
\ee
and the zeros or poles of $\tilde{\mu}$ are potentially the singular locus of the geometry. It is then straightforward to show that the EVH black hole singularity is  generically located at $\rho<0$ and it touches the horizon (which is at $\rho=0$) at some isolated points in $\theta$ where the zeros of $\mu_1$ are located.\footnote{As we will see in some specific examples this happens only at two points in $\theta$.} Away from these points the near horizon geometry of the EVH black hole is  expected to be smooth. This is how the EVH black holes are different from the so-called small black holes where the horizon and the singularity are basically becoming identical.

From above argument one can write the generic form of the metric after taking the EVH limit as follows
\be\label{generic-EVH-metric}
ds^2=-\rho \; \tilde{\mu} dt^2 + \frac{d\rho^2}{\rho^2 \tilde{\Lambda}} +\rho\; \tilde{g}_{\phi\phi} \left(d\phi+\tilde{N}^{\phi}dt\right)^2 + \tilde{g}_{\theta\theta}d\theta^2
\ee
where $\tilde{\mu}, ~ \tilde{\Lambda}, ~ \tilde{g}_{\phi\phi}, ~\tilde{N}^{\phi}$ and $\tilde{g}_{\theta\theta}$ are functions of $(\rho, \theta)$. In addition $\tilde{\mu}$ and $\tilde{\Lambda}$ do not have any zero in $[0,\infty)$.
Therefore, the most generic 4d EVH black hole metric is given by \eqref{generic-EVH-metric}, and is specified by five functions $\tilde\mu, \tilde\Lambda, \tilde g_{\phi\phi}, \tilde N^\phi$ and $\tilde g_{\theta\theta}$ which are all analytic functions of $\rho,\theta$.\footnote{The fact that \eqref{generic-EVH-metric} is  an extremal black hole with horizon at $\rho=0$, can also be manifestly seen noting that $g_{\rho\rho}^{-1}$ has double roots at $\rho=0$.   } These functions may be determined using the equations of motion \eqref{einstein}.
Finally, we note that
using the fact that the temperature (surface gravity) must be independent of the angular coordinate $\theta$, one can conclude that at the horizon (i.e. at $\rho=0$)
\be\label{mu-Lambda}
\tilde{\mu}(0,\theta)\tilde{\Lambda}(0,\theta)=L^2\,,
\ee
where $L$ is a  ($\theta$ independent) constant.

\section{Near horizon limit of EVH black holes}

One may now take the near horizon geometry of the EVH black hole \eqref{generic-EVH-metric}. This is done through the limit $\rho= \epsilon^2 r^2$, $\epsilon\rightarrow 0$. The resulting geometry is
\be\label{AdS3-solution}
ds^2= a(\tilde{\theta}) \left(-r^2 d\tilde{t}^2 + L^2  \frac{dr^2}{r^2} + b(\tilde{\theta}) r^2 d\tilde{\phi}^2 + R^2d\tilde{\theta}^2\right)
\ee
where $L, R$ are  constants and\footnote{To get the above metric we applied a coordinate transformation on $\theta$ to get the coefficient of $d\tilde{\theta}^2$ to be equal $a(\tilde{\theta})$ and, $\tilde{\omega}$ is the angular velocity of the horizon at EVH limit.}
\be\label{t-phi-scaling-EVH}
\tilde{t}={\epsilon} t\,,\qquad \tilde{\phi}={\epsilon}(\phi-\tilde{\omega} t)\,.\nonumber
\ee
We note that the above in particular implies that $\tilde\phi \in [0,2\pi\epsilon]$.

Besides the metric we also have gauge and scalar fields which we need to take care of. This will be done requiring \eqref{AdS3-solution} to be a solution to equations of motion of action \eqref{Ein-Dil-Max} with only $\tilde\theta$ dependent fields. This latter implies that all the components of the gauge field strength should vanish.\footnote{This is provided that $f(\Phi(\tilde\theta))\neq 0$ and is a positive definite function, which we assume it to be so.} In special case when the scalar potential is a constant $V=V_0$, if we take\footnote{One can always redefine scalar field to get the standard kinetic term in the action.} $G(\Phi)=1$ we can solve Einstein equations, from which  we learn that
\bea
&& \frac{d b}{d \tilde\theta}=0 \qquad  \Rightarrow \qquad b=b_0=const.\ ,\\  \nonumber
&& \frac{d^2 a}{d\tilde\theta^2}+\frac{4R^2}{L^2}a - V_0 R^2 a^2 = 0\,. \nonumber
\eea
The second equation can be rewritten in the following form
\be
\frac{d a}{d\tilde\theta}=\left(C-\frac{4R^2}{L^2}a^2 + \frac{2}{3}V_0R^2a^3 \right)^{1/2}  \nonumber
\ee
where $C$ is an integration constant. For $V_0=0$ case the above admits a simple solution
\be
\label{sola}
a=a_0\sin\frac{2R}{L}\tilde\theta\,,
\ee
where $C=\frac{2R}{L} a_0$.

The $\theta\theta$-component of the Einstein equations, which is compatible with the scalar field $\Phi$ equation of motion, yields
\be
\frac{d\Phi}{d\tilde{\theta}}= \pm\frac{\sqrt{3}C}{a}\,.   \nonumber
\ee
This  can be integrated easily when $V_0=0$, leading to
\be
\label{solphi}
\frac{d\Phi}{d\tilde{\theta}}= \pm \frac{\sqrt3 R}{L} \frac{1}{\sin(\frac{2R}{L}\tilde{\theta})}\ \qquad \Longrightarrow \qquad  e^{\frac{2\Phi}{\sqrt3}}= g_0 \tan \frac{R\tilde\theta}{L}\,,
\ee
where $g_0$ is a constant.

The above solution is specified by five parameters $a_0, b_0, R, L$ and $g_0$. Not all of these are physical.
$b_0$, as long as periodicity of $\tilde\phi$ direction is not specified, may be absorbed in the definition of $\tilde\phi$. Renaming $\frac{2R}{L}\tilde{\theta}$ as the coordinate $\theta$, $R$ dependence in the metric drops out and the geometry is conformal  to an \ads{3} times interval metric with the AdS$_3$ radius  given by $R_{AdS3}=a_0 L$.
Therefore, our final solution is specified by only two parameters, $g_0$ and the AdS$_3$ radius $R_{AdS3}$. Explicitly,
for the $V_0=0$ case, after solving the equations of motion, we end up with the solution
\be\label{AdS3-region}
\begin{split}
ds^2 &= R_{AdS3}^2 \sin\theta \left(-r^2 d{\tau}^2 +   \frac{dr^2}{r^2} +  r^2 d{\psi}^2 + \frac14 d{\theta}^2\right)\,,\\
F_{\mu\nu}&=0\,,\qquad e^{\frac{2\Phi}{\sqrt3}}= g_0 \tan \frac{\theta}{2}\,
\end{split}
\ee
where $\theta\in [0,\pi]$ and $\psi\in [0,2\pi\epsilon]$. We stress that the AdS$_3$ throat in \eqref{AdS3-region} in the near horizon limit of the EVH black hole  is a \emph{pinching} \ads{3}, because the circle inside
\ads{3}, $\psi$ has a vanishing periodicity  $2\pi{\epsilon}$. Note also that the above parameter count is true for any generic action, irrespective of $f(\Phi)$.

It is important to notice that, taking the near horizon limit and the EVH black hole limits (in the above specified respectively with $\tilde\epsilon$ and $\epsilon$ to zero limits) do not commute.
%%%%%%%%%%%%%%%%%%%%%%%%%%%%%%%%%%%%%%%%%%%%%%%%%%%%%%%%%%%%%%%%%%%%%%%%%%%%%%%%%%%%%%%%%%%%%%%%%%%%%%%%%%%5
%%%%%%%%%%%%%%%%%%%%%%%%%%%%%%%%%%%%%%%%%%%%%%%%%%%%%%%%%%%%%%%%%%%%%%%%%%%%%%%%%%%%%%%%%%%%%%%%%%%%%%%%%%%5
\subsection{Near horizon limit of near EVH black holes}
%%%%%%%%%%%%%%%%%%%%%%%%%%%%%%%%%%%%%%%%%%%%%%%%%%%%%%%%%%%%%%%%%%%%%%%%%%%%%%%%%%%%%%%%%%%%%%%%%%%%%%%%%%%5
%%%%%%%%%%%%%%%%%%%%%%%%%%%%%%%%%%%%%%%%%%%%%%%%%%%%%%%%%%%%%%%%%%%%%%%%%%%%%%%%%%%%%%%%%%%%%%%%%%%%%%%%%%%5

So far we have introduced the EVH black hole as a particular limit of a black hole solution with vanishing horizon area $A_h$ and temperature $T$.
For a general black hole solution this is not a smooth limit and we may get a naked singularity. However, as we have shortly discussed above and we will study in more details for the specific case of EVH KK black holes in section 4, the singularity of the EVH black hole is generically sitting behind the horizon which is at $\rho=0$, in the region with $\rho\leq 0$. The singularity, however, touches the horizon at two points in $\theta$ where $a(\theta)$ vanishes. At these two points we have naked singularity and away from these two points the geometry is smooth.

In this subsection we consider the near EVH black holes, black holes where $A_h$ and $T$ are non-zero but are small. Near EVH black holes, are hence defined by two parameters which specify how we have moved away from the EVH point. To study the near horizon limit of near EVH black holes, we start by the following double scaling limit
\be
\label{gdsl}
g_{\phi\phi}^{(0)}=\tilde{g}_{\phi\phi}^{(0)}\epsilon^2,\;\;\; \rho_+-\rho_-=\gamma^2 \epsilon^{v},
\ee
while taking the near horizon limit in the following way
\be
\rho=\rho_++\epsilon^{u} r^2\,.  \nonumber
\ee
As we will see momentarily, this will open up the possibility of defining near horizon, near EVH limit in a similar manner.
From (\ref{mu}-\ref{Lambda}) we get
\bea
\label{gnh1}
 ds^2 &=& - r^2\epsilon^u \left(\gamma^2 \epsilon^v+r^2 \epsilon^u\right)\mu(r,\theta) dt^2 + \frac{4 \epsilon^u dr^2}{ \left(\gamma^2 \epsilon^v+r^2\epsilon^u\right)\Lambda(r,\theta)} \cr\cr
&&  + \left(\tilde{g}_{\phi\phi}^{(0)}\epsilon^2 + \epsilon^u r^2 g_{\phi\phi}^{(1)} \right) (d\phi-\omega dt + \eta dt)^2 + g_{\theta\theta}^{(0)}d\theta^2\,.
\eea
As discussed earlier we want to keep $\Lambda$ finite therefore to get a regular geometry we need to restrict ourselves to $u \leq v$. For a given fixed value $v$ one can vary parameter $u$ to probe different regions of the near horizon.

\begin{figure}
 \begin{center}
 \includegraphics[scale=.45]{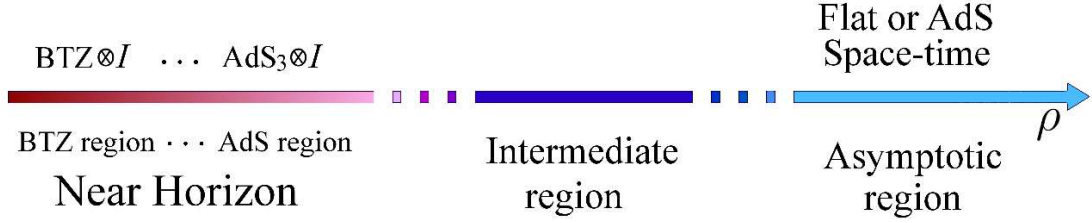}
 \end{center}
\caption{Near horizon structure of the EVH black hole in flat or AdS background. The Near horizon region has warped (\ads{3} or BTZ $\times I$) geometry. In the strict near horizon limit the intermediate and asymptotic regions are cut off from the geometry.}\label{NH-figure}
\end{figure}

It is clear from (\ref{gdsl}) that case $u=v$  corresponds to a case where the separation between the inner and outer horizons are distinguishable in after taking the near horizon limit. Let us call it BTZ region and the near horizon region corresponding to $u<v$, where the two horizons become essentially indistinguishable after the limit, the AdS region. This denomination will be evident soon. In Figure \ref{NH-figure} we have drawn the general structure of the near horizon geometry of an EVH black hole. From (\ref{gdsl}) we can consider ${\rm AdS}$ region as a far asymptotic region of the BTZ region.

We start with  $u < v$ case first. After dropping $\epsilon^v$ terms in (\ref{gnh1}) following the discussion in the previous section, if we desire to get a smooth solution with finite  $A_h/{T}$ ratio, we should take $u<2$ and require $\mu(r,\theta) \sim \frac{1}{\epsilon^{u}r^2}$ at near horizon. This is similar to what we discussed earlier in the previous section. After changing the coordinate and taking the limit, we get the geometry (\ref{AdS3-solution}) which upon imposing  the Einstein equations  we end up with the warped ${\rm AdS}_3 \times I$ solution \eqref{AdS3-region}, where $I$ is a line segment parameterising $\theta$ direction.

Let us now study the $u=v$ case, corresponding to the ``BTZ'' region. After a rescaling we have
\bea
ds^2 &=& - r^2(r^2+\gamma^2) \epsilon^{2u}  \mu(r,\theta)dt^2 + \frac{4dr^2}{(r^2+ \gamma^2)\Lambda(\theta)} + g_{\theta\theta}^{(0)}d\theta^2\cr &+& \left(\tilde{g}_{\phi\phi}^{(0)}\epsilon^2+\epsilon^u r^2 g_{\phi\phi}^{(1)}\right) (d\phi-\omega dt + \eta dt)^2
 \,.  \nonumber
\eea
In the above coordinate system,  location of inner and outer horizons of original black hole  solution correspond to $r^2=-\gamma^2$ and $r=0$,  respectively. Now, if we desire to get a black hole solution (rather than a naked singularity) we should assume that singularity is at $r^2<-\gamma^2$. As discussed, we expect a pole in $\mu$ at the location of singularity therefore at small $\epsilon$, function $\mu$ should behave as
\be
\mu = \frac{\mu_1(\theta)}{\epsilon^u (r^2+\alpha^2)}  \nonumber
\ee
where $\alpha^2> \gamma^2$ and $r^2=-\alpha^2$ is the location of the singularity.

We can see from (\ref{gnh1}) that the entropy scales as $A_h\sim \epsilon$, therefore to keep ratio $\frac{A_h}{T}$ finite we need to take $T\sim \epsilon$. On the other hand, (\ref{temperatur}) leads to
\be
T \sim \sqrt{\mu_1(\theta)\Lambda(\theta)} \epsilon^{u/2}\,.    \nonumber
\ee
In order to keep the geometry to be regular and $T\sim \epsilon$ we need to take $u=2$, and recalling
\eqref{mu-Lambda}, we end up with
\be
ds^2=-\frac{r^2(r^2+\gamma^2) \epsilon^2}{r^2+\alpha^2} \mu_1(\theta) dt^2 + L^2 \frac{ \mu_1(\theta) dr^2 }{ (r^2+\gamma^2)} + \epsilon^2 \left(\tilde{g}_{\phi\phi}^{(0)} +  r^2 g_{\phi\phi}^{(1)} \right) (d\phi-\omega dt + \eta dt)^2   + g_{\theta\theta}^{(0)}d\theta^2  \nonumber
\ee
After scaling and changing coordinates
\be
r^2\rightarrow r^2-\alpha\,,\qquad \tau=\epsilon t\,,\qquad \psi=\epsilon\phi\,.  \nonumber
\ee
we get
\bea
\label{NHG5}
ds^2=a(\theta)\bigg[- F(r) d\tau^2 + \frac{ dr^2}{F(r)} + \left(f(\theta)+ r^2 g(\theta)\right)  \left( d\psi+  h(r,\theta)d\tau\right)^2+\frac14d\theta^2 \bigg].
\eea
where
\be\label{F(r)}
F(r)=\frac{(r^2-r_+^2)(r^2-r_-^2)}{r^2}\,.
\ee

To specify the gauge fields one should  follow the limits over the gauge fields and try to solve Maxwell equations in above background. Besides the equations of motion we also need to determine the boundary conditions on different fields. To this end we  notice that  the region $v>u$  corresponds to the asymptotic (large $r$) region of  $u=v$ case. Therefore, we require all fields asymptotically to take values of the AdS region in \eqref{AdS3-region}. We observe that $g_{\tau\tau}$ and $g_{rr}$ components of the metric (\ref{NHG5}) have correct asymptotic behaviour. The boundary condition on $g_{\psi\psi}$ gives $g(\theta)=const.$ and also implies $h(r,\theta)$ falls down faster than $1/r$ at large $r$. Finally, we need to assume all gauge fields vanish asymptotically. By plugging the metric (\ref{NHG5}) along with  gauge  and scalar fields ansatz (\ref{GenSol}) into the equation of motion we obtain the equations for unknown functions $h(r,\theta), f(\theta)$ in metric and also the equations governing gauge  and scalar fields.

These equations are highly complicated partial differential equations and we are not able to solve them analytically. Instead we try to study them by a Taylor series expansion around the horizon. Assuming all functions appear in equations of motion have analytical expansion near the horizon of (\ref{NHG5}), which is located at $r=r_+$, we can expand them at the near horizon as follows
\be
X(r,\theta) = X^{(0)}(\theta) +  (r-r_+) X^{(1)}(\theta) + (r-r_+)^2 X^{(2)}(\theta) + \cdots,  \nonumber
\ee
where $X$ is any of unknown functions in the metric, gauge or scalar fields. By inserting these expansions into the equations of motion we get a series of algebraic and ordinary differential equations. When the potential of scalar field is a constant  we can solve the equations of motion in AdS region analytically which gives us \eqref{AdS3-region}. We are now able to solve analytically these series of algebraic and ordinary differential equations, to obtain  $F=0$ for all gauge fields, (\ref{solphi}) as the solution for scalar field and, the metric which takes the form
\be\label{BTZ-region}
ds^2 = R^2_{AdS3}\sin \theta\ \bigg[-F(r)d\tau^2 + \frac{ dr^2 }{F(r)} + r^2 (d\psi-\frac{r_+r_-}{r^2} d\tau)^2   +\frac14 d\theta^2 \bigg]\,,
\ee
where $F(r)$ is given in \eqref{F(r)}.
This is the same geometry as in \eqref{AdS3-region} but the \emph{pinching} \ads{3} has now been replaced with a \emph{pinching} BTZ geometry.
We will return to this latter point in section 5.

\subsection{EVH entropy function and EVH attractor mechanism?}
In previous subsections we discussed two near-horizon limits of a general (near) EVH black hole solution of 4d
Einstein gravity coupled with gauge fields and scalars. These are black holes with vanishing Hawking temperature $T$ and the horizon area $A_h$ while $A_h/T$ is fixed. In this section we study these near horizon geometries by using an extended notion of Sen's entropy function method \cite{Sen:2005wa}. A similar analysis has been carried out for charged asymptotically AdS$_4$ and AdS$_5$ EVH black holes \cite{Fareghbal:2008ar,Fareghbal:2008eh}.
In addition, using the entropy function method we compute the entropy of the  \emph{warped} ${\rm BTZ}\times I$ solution \eqref{BTZ-region} as the near-horizon limit of a rotating EVH black hole solution in our theory and show that this entropy is exactly equal to the entropy of the original 4d black hole solution.

The most general field configuration consistent with the local ${\rm SL}(2,\mathbb{C}) \times {\rm U}(1)$ symmetry of \emph{warped} ${\rm AdS}_3 \times I$ is of the form:
\bea
&& ds^2=v(\theta)^2\left(-\frac{(r^2-r_+^2)(r^2-r_-^2)}{r^2} dt^2+\frac{r^2dr^2}{(r^2-r_+^2)(r^2-r_-^2)}+r^2 (d\psi-\frac{r_+r_-}{r^2})^2+ R^2 d\theta^2\right),\crcr && F^I=0, \quad \Phi^A=u^A(\theta).  \nonumber
\eea
It is clear that the local ${\rm SL}(2,\mathbb{C}) \times {\rm U}(1)$ symmetry does not allow us to turn on the gauge fields. Next, we define the ``EVH entropy function'':
\be\label{EVH-ent-func}
f[u(\theta), v(\theta), R]= \int d\theta d\psi\, \sqrt{-g}\, {\mathcal L}
\ee
which is a function of $R$ and functional of $u(\theta)$ are $v(\theta)$, from which one can deduce the equations of motion
\be
\frac{\partial f}{\partial R}=0, \quad \frac{\partial f}{\partial u(\theta)}=0, \quad  \frac{\partial f}{\partial v(\theta)}=0,  \nonumber
\ee
and the entropy is given by
\be
S_{BH}= -f  \nonumber
\ee
where $f$ is evaluated at its extremum.\footnote{We would like to comment that the above notion of extended entropy function is closely related to the $c$-extremisation method of \cite{Kraus-Larsen}. Our EVH entropy function is nothing but the $c$-function there and in our case, due to the absence of gravitational Chern-Simons  we expect $c_L=c_R$ \cite{K-L-2}. To connect the two more closely, we note that one can reduce the action \eqref{Ein-Dil-Max} on the ansatz \eqref{EVH-ent-func} over the $\theta$ direction to obtain an effective 3d gravity theory. Then, one may recall the Cardy formula and that, for a 2d CFT (or BTZ black hole) at a fixed temperature the (Wald) entropy and central charge are proportional to each other \cite{Jiro-us}.}

For the Lagrangian (\ref{Ein-Dil-Max}) with vanishing scalar potential the EVH entropy function is given by
\be
f=\frac{2\pi\epsilon r_+  }{\kappa_4^2} \int d\theta \left(6R v^2 -\frac{6v'^2}{R} +\frac{2 G_{AB }(u) u'^Au'^B v^2}{R} \right),  \nonumber
\ee
where prime is the derivative with respect to $\theta$. The equations of motion are obtained as
\bea
\frac{\partial f}{\partial R}=0 \quad&\Rightarrow& \quad 3v^2+\frac{3v'^2}{R^2}-\frac{G_{AB }(u) u'^Au'^B v^2}{R^2}=0 \nonumber \\
\frac{\partial f}{\partial v}=0 \quad&\Rightarrow& \quad3 Rv + \frac{3 v''}{R} + \frac{G_{AB }(u) u'^Au'^B v}{R}=0,\nonumber \\
\frac{\partial f}{\partial u^C}=0 \quad&\Rightarrow& \quad\frac{d G_{AB}}{du^C} u'^Au'^B v^2 -2(G_{AC}(u) u'^C \nonumber v^2)'=0.
\eea

In the case of a single scalar field if we take $G=1$, one can simply solve above equations to obtain \eqref{AdS3-region} or \eqref{BTZ-region}. Finally the entropy is evaluated as follows
\be\label{near-EVH-entropy}
S_{BH}=\frac{2 \pi R^2_{AdS3}}{\kappa_4^2} \cdot(2\pi \epsilon r_+)=\frac{\pi\epsilon r_+}4\cdot \frac{R^2_{AdS3}}{ G_4}\,.
\ee
Note that the factor of $2\pi \epsilon$ has appeared from the integration over $\psi$ direction.
The above equation has an interesting and simple interpretation: In the EVH limit, although the horizon topology of the near EVH black hole is still a two-sphere, its geometry is very close to a thin cylinder, the axis of which is along the $\theta$ direction and its hight is $2R_{AdS3}$, while its circle spanned by the $\psi$ direction, is a circle of radius $R_{AdS3} r_+\epsilon$. In the EVH limit, one cannot essentially distinguish the spherical topology from a cylindrical one. We will return to this point in section 4.1 and also in the discussion section.

One of the interesting and important features of  extremal black holes is the attractor behaviour \cite{attractor,{Sen:2005wa}}: the value of moduli, and in general the whole geometry, on the horizon are only specified by the charges and are independent of their values at asymptotic space-like infinity. As a result the entropy is only a function of (quantised) charges and not the value of moduli. This fact has been made manifest in the entropy function formalism \cite{Sen-AdS2/CFT1}. \footnote{If there are flat directions for some of the moduli their value is not fixed by the charges at the horizon, nonetheless these moduli do not contribute to the entropy either.} In the EVH case, the entropy for the near EVH geometry \eqref{near-EVH-entropy}, depends on two parameters, $r_+$ and $R_{AdS3}$, none of which are related to the value U(1) gauge field charges at the horizon. However, as we see in explicit examples of the next section and although not explicitly  from the above near horizon geometry construction, the \ads{3} radius $R_{AdS3}$,  is specified by the charges of the black hole at infinity, while $r_+$ remains a free parameter.

We also comment that our near EVH geometries are not necessarily extremal. Following the same reasoning as we have presented in section 3.1, it is possible to show that if we start with an extremal near EVH black hole its near horizon geometry will contain an extremal (pinching) BTZ factor. This will be made explicit in the  KK black hole example of next section.

\section{Some specific examples}
In this section we choose two specific classes of 4d black holes containing  EVH black holes, one in the class of rotating charged asymptotically flat black holes and the other in asymptotically \ads{4} static charged black holes of the $U(1)^4$ gauged supergravity. These are two special cases of the EMD theories \eqref{Ein-Dil-Max}. We customise the general arguments of section 3 for these cases and study their near horizon geometry in further detail.

%%%%%%%%%%%%%%%%%%%%%%%%%%%%%%%%%%%%%%%%%%%%%%%%%%%%
\subsection{Rotating Kaluza-Klein EVH black holes}
%%%%%%%%%%%%%%%%%%%%%%%%%%%%%%%%%%%%%%%%%%%%%%%%%%%%
The most general KK black holes constitute a four parameter black hole family which are solutions to the 4d gravity theory:
\be
\label{action0}
I = \frac{1}{16 \pi G_4} \int d^4x \sqrt{-G_{(4)}}\left( R_{(4)}-2\partial^{\mu}\Phi \partial_{\mu}\Phi- e^{2\sqrt{3}\Phi}F^{\mu\nu}F_{\mu\nu}\right)\,.
\ee
This action is obtained from the KK-reduction of 5d Einstein gravity to four dimensions and as such each solution to 4d theory \eqref{action0} has a five-dimensional uplift of the form \cite{KK-black-hole-1}
\be
\label{5dsolution}
ds_{(5)}^2= e^{4\Phi/\sqrt{3}} (dy +2 A_{\mu} dx^{\mu})^2 + e^{-2 \Phi/\sqrt{3}} ds_{(4)}^2\,.\nonumber
\ee
For the KK black holes of our interest the 4d metric is \cite{KK-black-hole-1}
\be
\label{4d-metric}
ds^{2}_{(4)} = -\frac{\tilde{\Delta}}{\sqrt{f_pf_q}} (dt- \omega )^2 +\frac{\sqrt{f_pf_q}}{\Delta} d\rho^2 + \sqrt{f_pf_q} d\theta^2 + \frac{\Delta \sqrt{f_pf_q}}{\tilde{\Delta}} \sin^2\theta d\phi^2 \nonumber
\ee
where
\bea
\label{fp}
f_p &=& \rho^{2}+m^{2}j^2\cos^{2}\theta+\rho(p-2m)+{p\over
p+q}{(p-2m)(q-2m)\over 2} \nonumber  \\ &~&\qquad - {p\over 2(p+q)}
\sqrt{(p^{2}-4m^{2})(q^{2}-4m^{2})}~j\cos\theta~,\nonumber  \\
\label{fq}
f_q &=& \rho^{2}+m^{2}j^2\cos^{2}\theta+\rho(q-2m)+{q\over
p+q}{(p-2m)(q-2m)\over 2}  \nonumber \\ &~&\qquad +
{q\over 2(p+q)}
\sqrt{(p^{2}-4m^{2})(q^{2}-4m^{2})}~j\cos\theta~,\nonumber \\
\tilde{\Delta} &=& \rho^{2}+m^{2}j^2\cos^{2}\theta-2m\rho~,\nonumber \\
\Delta &=& \rho^{2}+m^{2}j^2-2m\rho~, \nonumber\\
\omega &=& \sqrt{pq}{(pq+4m^{2})\rho-m(p-2m)(q-2m)\over 2(p+q)\tilde{\Delta}}~j
\sin^{2}\theta d\phi~,\nonumber\\
A&=&-f_{q}^{-1}\left({Q\over 4\sqrt\pi}\, \left(r+
\frac{p -2m}{2}\right)+\frac{1}{2}
j\sqrt{\frac{q^{3}
\left(p^{2}-4m^{2}\right)}{4\left(p +q\right)}}\cos\theta\right) dt
\nonumber \\
&& -\bigg( {P\over 4\sqrt\pi}\, \cos\theta +f_{q}^{-1}
{P\over 4\sqrt\pi}\, m^2j^2\sin^{2}\theta\cos\theta
 \nonumber \\
 &  & +\frac{1}{2}f_{q}^{-1}\sin^{2}\theta
 j\sqrt{\frac{p \left(q^{2}
 -4m^2\right)}{4\left(p +q\right)^{3}}}
 \left[ (p +q)(p r-m(p -2m))+
 q(p^{2}-4m^2)\right]\bigg) d\phi\,\nonumber,
\eea
and the solution for the dilaton is of the form
\be
e^{-\frac{4}{\sqrt{3}}\Phi}=\frac{f_p}{f_q}\nonumber \,.
\ee

The four parameters $(m,j,q,p)$ appearing in the solution are related to the (four-dimensional)
physical parameters, mass $M$, angular momentum $J$, electric charge $Q$, and magnetic
charge $P$ as
\bea
M&=& 4\pi(p+q)~,
\label{eq:paraM} \nonumber\\
G_4 J &=& {\sqrt{pq}(pq+4m^2)\over 4(p+q)}~j~,
\label{eq:paraJ} \nonumber\\
Q^{2} &=& 4\pi {q(q^2-4m^2)\over (p+q)}~,
\label{eq:paraQ} \nonumber\\
P^{2} &=& 4\pi {p(p^2-4 m^2)\over (p+q)}~.
\label{eq:paraP}\nonumber
\eea
Therefore, non-nakedly-singular black hole solutions should necessarily have $q\geq 2m$, $p\geq 2m$, $j\leq 1$. Moreover,  regularity implies that the coordinate $y$ must be periodically identified as
\beq\label{yperiod}
y\sim y+ 2\pi R\,,\qquad R=\frac{P}{\sqrt{\pi}N_6}\,,\nonumber
\eeq
for integer $N_6$. In five-dimensions the electric charge is the momentum
along the $y$-direction and  is hence quantised as
\begin{equation}\label{eq:quant_Q}
Q=\frac{8 \sqrt{\pi}G_4N_{0}}{R}\, \nonumber
\end{equation}
for integer $N_0$. The KK black hole may be embedded in the string theory to a rotating D0-D6 brane system \cite{KK-black-hole-2,{KK-black-hole-3}} where $N_0$ and $N_6$
correspond to the numbers of D0 and D6 branes and $R=g$ in string units.
In five-dimensions, when the black hole size (with a non-zero magnetic charge) is much smaller than
the compact radius, the above can be considered as black holes sitting at the tip of a
Taub-NUT space \cite{KK-black-hole-3,{Sen:2005wa}}.

We note that the action \eqref{action0} has a scaling symmetry
\be
\Phi \rightarrow \Phi+\Phi_{\infty}, \quad  F_{\mu\nu} \rightarrow  {\rm e}^{-\sqrt{3}\Phi_{\infty}} F_{\mu\nu} , \nonumber
\ee
for a constant $\Phi_{\infty}$. Therefore, we can generate one parameter family of solutions carrying
fixed electric and magnetic charges by using the transformation
\be
\Phi \rightarrow \Phi+\Phi_{\infty}, \quad  F_{\mu\nu} \rightarrow  {\rm e}^{-\sqrt{3}\Phi_{\infty}} F_{\mu\nu}
 \quad Q \rightarrow  {\rm e}^{-\sqrt{3}\Phi_{\infty}} Q \quad P \rightarrow  {\rm e}^{\sqrt{3}\Phi_{\infty}} P\,.  \nonumber
\ee

\subsubsection{EVH KK black holes}
Generic KK black holes are stationary geometries.  A rotating KK black hole can be in either of  fast-rotating ($G_4 J> PQ$) or  slowly rotating
($G_4J<PQ$) branches \cite{KK-black-hole-2}. In the extremal cases, the Kerr/CFT-type analysis has been carried out for both cases and shown that in these cases there are different U(1) isometries which enhance to chiral Virasoro of the proposed dual chiral 2d CFT \cite{Azeyanagi:2008kb}. Here, however, we are interested in the overlap of the two branches, i.e. $PQ=G_4J$, where we have an EVH KK black hole. This geometry is often dismissed being a naked singularity \cite{KK-black-hole-2}. We will comment on this point below.

To find the EVH point, we recall that the temperature and horizon area of the KK black hole are given by%
\be\label{T-S}%
T = \frac{m}{\pi \sqrt{pq}} \left[ \frac{pq +4m^2}{p+q} +
\frac{2m}{\sqrt{1-j^2}} \right]^{-1},\,\,
{A_h} = 2\pi{\sqrt{pq}}\sqrt{ 1-j^2} \left[
{pq+4m^2\over (p+q)} +\frac{2m}{\sqrt{ 1-j^2}}\right].
\ee%
It is seen that in the  limit $m\to 0$, $j\to 1$ with
$\sqrt{1-j^2}/m$ held fixed, which may be achieved by  the following scaling%
\be\label{Near-EVH}%
m=\epsilon\mu,\ 1-j^2=\epsilon^2\lambda^2\,,\qquad \epsilon\to 0\,,
\ee
the area to temperature ratio remains fixed. In other words, in the four dimensional ($n=4$) parameter space of  rotating KK black holes, there exists a two dimensional EVH hypersurface parameterised by $p,q$.

The EVH KK black hole metric is then obtained by setting $m=0,\ j=1$, for which the metric takes the form
\be
\label{EVHKK}
ds^2=-N^2 dt^2 + g_{\rho\rho}d\rho^2 + g_{\phi\phi} \left(d\phi+N^{\phi}dt\right)^2 + g_{\theta\theta}d\theta^2
\ee
where functions $N^2, g_{\rho\rho}, g_{\phi\phi}, N^{\phi}$ and $g_{\theta\theta}$ are given by
\bea
&& g_{\rho\rho}= \frac{\sqrt{f_{pq}}}{\rho^2}, \;\;\;\; g_{\theta\theta}= {\sqrt{f_{pq}}} \cr\cr
&& g_{\phi\phi} = \rho\;\frac{\sin^2\theta}{\sqrt{f_{pq}}}\left(\rho(\rho+p)(\rho+q)+ \frac{pq }{2}\rho \;(1+\frac{q-p}{p+q}\cos\theta
)+\frac{p^2q^2}{p+q}\right)\cr\cr
&& N^{\phi} = -\frac{\;\; p^{3/2}q^{3/2}}{2(p+q)} \left( \rho(\rho+p)(\rho+q)+ \frac{pq }{2}\rho \;(1+\frac{q-p}{p+q}\cos\theta
)+\frac{p^2q^2}{p+q}\right)^{-1}\cr\cr
&& N^2= \frac{\rho}{\sqrt{f_{pq}}}\left[\rho+ \frac{p^3q^3\sin^2\theta}{4(p+q)^2} \left( \rho(\rho+p)(\rho+q)+ \frac{pq }{2}\rho \;(1+\frac{q-p}{p+q}\cos\theta
)+\frac{p^2q^2}{p+q}\right)^{-1}\right]  \nonumber
\eea
and function $f_{pq}$ is defined by
\be
\label{fpq}
f_{pq}=\left(\rho^2(\rho+p)(\rho+q)+ \frac{pq}{2}\rho^2 \;(1+\frac{q-p}{p+q}\cos\theta
)+\frac{p^2q^2}{p+q}\rho + \frac{p^3q^3}{4(p+q)^2}\sin^2\theta \right)\,.
\ee
Scalar and gauge fields are given by
\bea
&& A_t= -\frac{q^{3/2} \sqrt{p+q}\left[2 \rho+ p(1+\cos\theta)\right]}{2\rho(\rho+q)(p+q)+q^2p(1+\cos\theta)}\,,\nonumber \\ \cr\cr \nonumber
&& A_{\phi}= -\frac{p^{3/2} \left[2\rho(\rho+q)(p+q)+q^2p(1+\cos\theta)+\rho q(p+q)\sin^2\theta\right]}
{\sqrt{p+q} \left[2\rho(\rho+q)(p+q)+q^2p(1+\cos\theta)\right]}\,, \nonumber \\ \cr\cr
&& e^{\frac{4\Phi}{\sqrt{3}}}= \frac{2\rho(\rho+q)(p+q)+q^2p(1-\cos\theta)}{2\rho(\rho+p)(p+q)+qp^2(1+\cos\theta)}\,.  \nonumber
\eea

The horizon of the above EVH KK black hole is located at $\rho=0$. The location of singularity is given by zeros of $f_{pq}$:
$$\rho_s=\rho_s(\theta)\,,\qquad f_{pq}(\rho_s,\theta)=0\,.
$$
From (\ref{fpq}) we observe that for generic values of $\theta$ the singular line lies in the  $\rho<0$ region. More precisely, $\rho_s\leq 0$ and the equality happens for $\theta=0$ and $\theta=\pi$, that is $\rho_s(0)=\rho_s(\pi)=0$. This is the picture we discussed before. The singularity which is located at negative $\rho$ is generically sitting behind the horizon  which is at $\rho=0$. The singularity becomes ``naked'' only in two points: $\rho=0, \theta=0$ and $\rho=0, \theta=\pi$. For the EVH KK black hole, therefore, away from these two singular points, horizon is generically far from singularity, and indeed we define our near horizon limit such that, generically, we are parametrically infinitely far from the singularity.
This is how the EVH KK black hole is different from the ``small black holes'' where the singularity and horizon are always arbitrarily close.\footnote{For small black holes in string theory it has been shown that \cite{Sen:2004dp} adding the higher derivative corrections blows up the horizon to non-zero size and the resulting Bekenstein-Hawking entropy precisely matches with counting the corresponding microstates. It is interesting to study the effect of higher derivative corrections to the horizon shape for EVH black holes.}
Moreover, from the above metric one can find the geometric shape of the horizon of the EVH KK black hole. This is topologically a two-sphere, but a singular one, because $g_{\phi\phi}$ vanishes at the horizon $\rho=0$. In other words, close to the horizon and at constant $\rho, \tau$, the  metric is more like a cylinder the axis of which is along $\theta$ direction and its circle, which has vanishing radius is along $\phi$ direction.

\begin{figure}
\label{horizon}
 \begin{center}
 \includegraphics[scale=.65]{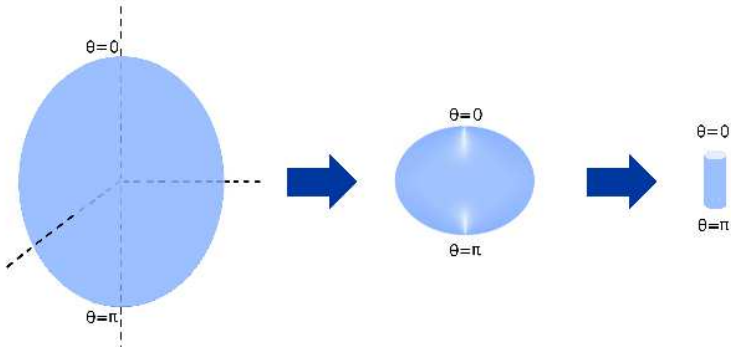}
 \end{center}
\caption{Horizon geometry as we increase angular momentum (left to right). The left figure shows the horizon at $J=0$.  For a fixed $P,Q$, increasing angular momentum reduces the horizon area and at the critical value $G_4J=PQ$, the EVH point, horizon area vanishes. For the near EVH, i.e when $|G_4J-PQ|\ll PQ$, horizon has the shape depicted in the right figure: it is a thin cylinder.}\label{horizon}
\end{figure}

\subsubsection{Near horizon limit of EVH KK black hole}
One may study the near horizon limit of the geometry obtained in the
(near) EVH limit. In order this, let us consider (\ref{Near-EVH})
and apply the scaling
\be\label{NH-EVH-coord-scaling}%
\rho=\epsilon^2\frac{pq}{p+q}\ r^2,\qquad
t=\frac{2\sqrt{pq}}{{\epsilon}}\tau,\qquad
\phi-\frac{t}{2\sqrt{pq}}=\frac{1}{{\epsilon}}\ \psi\ , %
\ee%
with $r,\ \tau$ and $\psi$ held fixed. In this limit, the 4d metric
\eqref{Near-EVH} takes the form
\be\label{EVH-NH-metric}%
ds^2 =R_{AdS3}^2 |\sin\theta|\left[-r^2 d\tau^2+\frac{dr^2}{r^2}+r^2
d\psi^2+\frac{1}{4} d\theta^2\right]\ ,
\ee%
where%
\be\label{AdS3-radius}%
R^2_{AdS3}=8PQ=\frac{2(pq)^{3/2}}{p+q}\,.%
\ee%
Gauge field and scalar potential in this limit
are
\be \label{NHEVH-F-Phi}
A_{t}=-\sqrt{\frac{p+q}{4q}}, \;\;\;\;\;A_{\phi}=\sqrt{\frac{p^3}{p+q}}\,,\qquad e^{\frac{4\phi}{\sqrt{3}}} = \frac{q}{p}
\tan^2\frac{\theta}{2}
\ee
and therefore gauge field strength vanishes.

As we see the near horizon limit \eqref{NH-EVH-coord-scaling} and the near horizon metric \eqref{EVH-NH-metric} are  exactly of the form that were outlined and discussed in the previous section. In this case, however, the \ads{3} radius $R_{AdS3}$ and the value of the dilaton field $g_0$ are determined by the value of the charges $p,q$ defining the EVH KK black hole. Although not implied by the equations of motion on the near horizon geometry  (\emph{cf.} discussions of section 3), the value of these two parameters are fixed by the charges defining the full EVH KK black hole, once it is extended out of horizon and to asymptotic flat region. In this sense, the EVH KK black hole still shows ``attractor behaviour''.

\subsubsection{Near horizon limit of near EVH KK black hole}

One may follow the steps of section 3.1 and study the near horizon limit of \emph{near}{ EVH} KK black hole. To this end, let us consider
\eqref{Near-EVH} together with the following scaling
\bea
&& \label{rscaling}\rho=\frac{pq}{p+q}\; \left(r^2-{r_+^2}\right)\; \epsilon ^2 + \rho_{+} \\
&& \label{tscaling} t = 2\sqrt{pq} \;\frac{\tau}{{\epsilon}} \\
&& \label{phiscaling}\phi = \frac{\tau+ \psi}{{\epsilon}}
\eea
where $r_{\pm}$ are given by
\be\label{r-pm}
r_{\pm} = \frac{2\mu (p+q) \pm pq\lambda}{2pq}\,.
\ee
Taking the limit $\epsilon \rightarrow 0$, we obtain the following geometry
\bea\label{NH-NEVH-metric}
ds^2= R_{AdS3}^2|\sin\theta| \bigg( -F(r)d\tau^2 + \frac{ dr^2}{F(r)} + r^2 (d\psi-\frac{r_+ r_-}{r^2}d\tau)^2 + \frac{1}{4} d\theta^2\bigg)
\eea
where $$F(r)=\frac{(r^2-r_+^2)(r^2-r_-^2)}{r^2},$$
$\theta\in [0,\pi]$ and, the gauge field and dilaton take the same values as in \eqref{NHEVH-F-Phi}. That is, the pinching \ads{3} in \eqref{EVH-NH-metric} is now replaced by a pinching BTZ, again in accord with our general discussions of section 3.1.

We note that taking the above near horizon near EVH limit do not change the entropy. This will have implications for the discussions in section 5. To see the latter, one may reduce the 4d gravity theory \eqref{action0} over the metric ansatz
\be
ds^2=R^2_{AdS3}|\sin\theta|\left( g_{ab} dx^a dx^b+\frac14 d\theta^2\right)\,,  \nonumber
\ee
with $a,b=1,2,3$ and $\theta\in [0,\pi]$. In the gravity sector of the 3d reduced action we obtain an \ads{3} theory with 3d cosmological constant $-R_{AdS3}^{-2}$ and 3d Newton constant
\be\label{G3-G4}
G_3=\frac{2G_4}{R_{AdS3}}\,.
\ee
The Bekenstein-Hawking entropy of the pinching BTZ solution to this 3d theory is then%
\be\label{3d-entropy}
S_{3d}=\frac{2\pi\epsilon r_+ R_{AdS3}}{4G_3}=\frac{R^2_{AdS3}}{G_4}\cdot \frac{\pi \epsilon r_+}{4}\,.
\ee
The entropy of  the original KK black hole (\emph{cf.} \eqref{T-S}) is given by%
\be\label{4d-entropy}
S_{4d}=\frac{A_h}{4G_4}= \frac{2\pi \sqrt{pq}\ \lambda  \epsilon}{4G_4}\left( \frac{pq}{p+q}+\frac{2\mu}{\lambda}\right)=\frac{R^2_{AdS3}}{G_4}\cdot \frac{\pi \epsilon r_+}{4}\,,
\ee
where  we have used \eqref{Near-EVH}, \eqref{AdS3-radius} and \eqref{r-pm}. We also note that the above result is in agreement with our general discussions leading to \eqref{near-EVH-entropy}. It is also instructive to compare the Hawking temperatures of the original near EVH KK black hole and that of the pinching BTZ:
\be\label{4d-temperature}
T_{4d}=\frac{\epsilon}{2\sqrt{pq}}\ T_{BTZ} \;
\ee
where $T_{BTZ}= \frac{r_+^2-r_-^2}{2\pi r_+ }$. The prefactor $\frac{\epsilon}{2\sqrt{pq}}$ is expected recalling \eqref{tscaling}.

Being a solution to 4d KK theory, the near horizon EVH black hole solution may  simply be uplifted to 5d:\footnote{We note that at the  singular point $\theta=0$ the coupling $e^{2\Phi/\sqrt{3}}$ is small and hence the 4d description is a good one. In the other singular point $\theta=\pi$ the coupling blows up \eqref{NHEVH-F-Phi} and one should use the 5d uplift.}%
\be\label{5d-uplift}%
ds^2_5= 2\sqrt{\frac{p}{q}}\left\{\cos^2\vartheta\left[-r^2
d\tau^2+R^2_{AdS3}\frac{dr^2}{r^2}+r^2
d\psi^2\right]+R^2_{AdS3}\left[\cos^2\vartheta
d\vartheta^2+\tan^2\vartheta\ d\chi^2\right]\right\}\ %\nonumber
, \ee
where $\vartheta=\theta/2$, $0\leq \vartheta\leq \pi/2$ and %
\be
4P\chi=y-\sqrt{\frac{p+q}{q}}t+p\sqrt{\frac{p}{p+q}} \varphi\ .\nonumber
\ee
We have normalised $\chi$ such that it is a periodic variable
ranging over $[0,2\pi]$. This 5d geometry is exactly the one obtained by Bardeen and Horowitz \cite{bardeenhorowitz} as the near horizon limit of EVH 5d Kerr solution. This is expected since the EVH KK black hole  is the 5d EVH Kerr at
the tip of the NUT.\footnote{We would like to thank Roberto Emparan for the comment on this point.}
As expected the process of uplifting does not change temperature or entropy. We also comment that along the discussions of \cite{KK-black-hole-2,Emparan-I} this solution may be embedded in string theory.

\subsection{Static charged black holes in 4d $U(1)^4$ SUGRA}

The second example of the 4d EVH black holes we discuss briefly here is within the family of electrically charged
\emph{static} black hole solution $U(1)^4$ SUGRA. The solution is specified by five parameters, four charges $q_I$ and $\mu$, and  is given by \cite{AdS4-general-BH}
\bea
&& ds^2=-H^{-1/2}f dt^2 + H^{1/2}\left(\frac{dr^2}{f}+r^2d\Omega_2^2\right), \nonumber\\
&& A_I=\frac{\tilde{q}_I}{q_I}\left(\frac{1}{H_I}-1\right), \;\;\;\; X_I=\frac{H^{1/4}}{H_I},  \nonumber
\eea
where
\bea
H_I=1+\frac{q_I}{r},\;\;\;\; H=H_1H_2H_3H_4,\;\;\;\; f=1-\frac{\mu}{r}+\frac{4r^2}{L^2}H,\;\;\;\;\nonumber \tilde{q}_I=\sqrt{q_I(q_I+\mu)}\,.
\eea

For the three-charge case, e.g. $q_1=0$ case, and with  $\mu=\frac{4q_2q_3q_4}{L^2}$ we obtain an EVH black hole. One may study its  near horizon near EVH limit. This has been studied in detail in \cite{Fareghbal:2008eh} and we just quote the main results and features here.
\begin{itemize}
\item In the family of three-charge EVH black holes we have both supersymmetric EVH and non-supersymmetric EVH
    solutions.
\item Although the 4d (EVH) black hole solution is static, its 11d uplift, where the geometry corresponds to three stacks of intersecting  rotating (giant) spherical M5-branes \cite{AdS4-general-BH}, is stationary.
\item The near horizon EVH limit is singular in 4d, while the limit is well-defined over the 11d uplift of the solution.
\item  In this 11d description the near horizon geometry of supersymmetric EVH geometry develops an \ads{3} throat \emph{without the pinching} issue, while for the non-supersymmetric case we obtain a \emph{pinching} \ads{3} throat in the same way we described in section 3. In both cases the ``circular'' part of \ads{3}, the $\psi$ direction in the notation of previous sections, comes from the seven dimensional part of the 11d solution and is not a part of the original asymptotic \ads{4} geometry. The $r,t$ part of \ads{3} throat, on the other hand, is coming from the $t$ and $r$ directions of original 4d black hole.

\item The above EVH solutions interpolate between the \ads{3} throat on the horizon and  \ads{4} in the asymptotic (large $r$) region. The radius of the \ads{3} is specified by the three charges $q_2,q_3,q_4$ \cite{Fareghbal:2008eh}.

\item In both the supersymmetric and non-supersymmetric cases, the entropy of the near EVH geometry before taking the near horizon limit is equal to that of the 3d BTZ geometry, in the same manner discussed in section 3.2 and for the KK black hole case in section 4.1.
\end{itemize}

\section{ EVH/CFT correspondence}

So far we have studied some near horizon limits of (near) EVH black holes, showing that we will generically obtain an \ads{3} throat. We also showed that the entropy of the original near EVH black hole is parametrically equal to the entropy of BTZ geometry obtained in the near horizon limit. In this section we first show that this near horizon limit is indeed a decoupling limit, in the same sense as in the standard AdS/CFT. Based on these observations and arguments, we will then propose the EVH/CFT correspondence, describing physics on the background of near horizon EVH black holes in terms of a dual 2d CFT.

\subsection{Near horizon limit as a decoupling limit}

To argue for the decoupling of low energy physics in near horizon EVH geometry  we consider equation of motion of a massless scalar field in the background
of EVH-KK black hole (\ref{EVHKK}), i.e. the geometry before taking the near horizon limit:
\be
\frac{1}{\sqrt{-g}} \partial_{\mu}\left(\sqrt{-g} g^{\mu\nu} \partial_{\nu}\Psi\right)=0\,.  \nonumber
\ee
Using metric (\ref{EVHKK}) the above equation reads
\bea \label{waveq}
&& -\frac{\sqrt{-g} }{N^2} \partial_t^2 \Psi + \sqrt{-g} \left(\frac{N^2-g_{\phi\phi}(N^\phi)^2}{N^2}\right) \partial_{\phi}^2\Psi + 2 \partial_{t \phi}\left(\sqrt{-g}\ \frac{N^\phi}{N^2} \Psi\right)
\cr\cr && +
\partial_{\rho}\left(\sqrt{-g} \frac{1}{g_{\rho\rho}} \partial_{\rho} \Psi\right)
+
\partial_{\theta}\left(\sqrt{-g} \frac{1}{g_{\theta\theta}} \partial_{\theta} \Psi\right)=0\,.
\eea

To solve the equation let us consider an ansatz for the partial wave solution with the following form
\be\label{Psi-ansatz}
\Psi=e^{-i(\omega t + k\phi)} F_{lk\omega}(\rho) Y_{lk\omega}(\theta).
\ee
Plugging this ansatz into equation (\ref{waveq}), the $\theta$ dependent part of equation takes the form
\be\label{diffeqY}
\ddot{Y}_{lk\omega}-\frac{2x}{1-x^2} \dot{Y}_{lk\omega} + \frac{P_{lk\omega}(x)}{(1-x^2)^2}Y_{lk\omega} =0,
\ee
where $x=\cos\theta$ and $dot$ denotes derivative with respect $x$ and the potential $P$ is the following polynomial of $x$
\bea
P_{lk\omega}(x)= l(l+1)-k^2-\frac{pq\omega^2(p-q)}{2(p+q)}x - l(l+1)x^2 + \frac{pq\omega^2(p-q)}{2(p+q)}x^3. \nonumber
\eea
For special case when electric and magnetic charges are equal, i.e. for  $p=q$, the solution of differential equation (\ref{diffeqY}) is simply expressed as associated Legendre polynomials
\be
{Y}_{lk\omega}= P_l^k(x)\equiv \frac{(-1)^k}{2^l l!}(1-x^2)^{k/2}\frac{d^{l+k}}{dx^{l+k}}(x^2-1)^l\,.  \nonumber
\ee
For the general $p\neq q$ case, if $Y=(x^2-1)^{k/2} H$, then $H$ satisfies the confluent Heun's equation, see the Appendix for the details.

The radial equation for $F_{lk\omega}$, recalling  the near horizon scaling \eqref{NH-EVH-coord-scaling},  is more conveniently written  in terms of $r$ variable, $\rho=r^2$. Defining the new function $R$,
\be
F_{lk\omega}=\frac{R_{lk\omega}(r)}{r^{3/2}}\, % \nonumber
\ee
the radial equation takes  the form of a ``Schr\"{o}dinger equation''
\be
R''_{lk\omega}-U(r) R_{lk\omega}=0\,, % \nonumber
\ee
where $prime$ denotes derivative with respect $r$ and potential $U$ is given by
\be\label{shrod-potential}
U(r)=-4\omega^2r^2-4{(p+q)\omega^2}+\frac{4l(l+1)+3/4-6pq\omega^2}{r^2}-\frac{4\omega p^{3/2}q^{3/2}(k+\sqrt{pq}\omega)}{(p+q)r^4}\,.
\ee

\begin{figure}
\label{potential}
 \begin{center}
 \includegraphics[scale=.45]{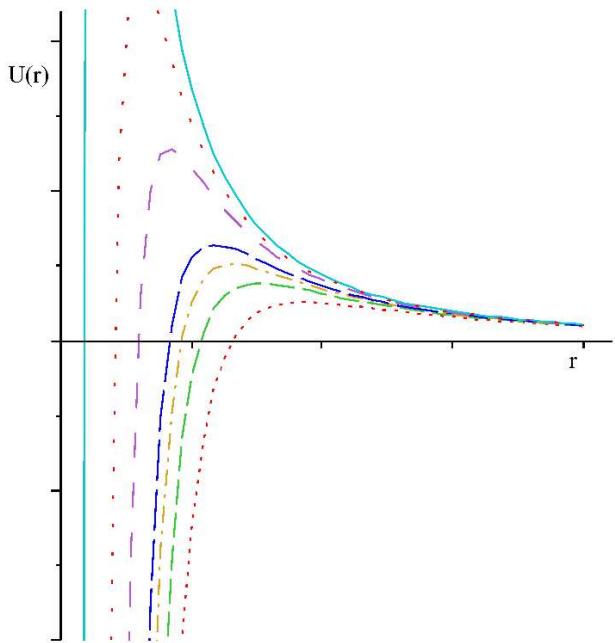}
 \end{center}
\caption{Behaviour of the potential as we vary $\omega$, and for $\omega(k+\sqrt{pq}\omega)>0$. As we decrease $\omega$ the hight of the maximum of the potential  increases and in the low energy $\omega\to 0$ limit the potential develops an infinite barrier, signaling the decoupling.}
\end{figure}

We are interested in the behaviour of the potential around $r=0$, where the horizon of the EVH black hole resides and in low energy, when the energy and angular momentum of the probing particle $\omega, k$ is scaling to zero (\emph{cf.}\eqref{tscaling}, \eqref{phiscaling}). If the coefficient  of the dominant term, the $r^{-4}$ term, is negative, that is when  $\omega(k+\sqrt{pq}\omega)>0$,  then the potential has a maximum around $r^2_{max}\propto \omega$ and the value of the potential at this maximum scales like $1/\omega$, which blows up in the limit. Therefore, an infinite potential barrier develops
which decouples the near horizon dynamics from the rest of the space.\footnote{We comment that the $r$ coordinate is a suitable one for exploring the near horizon small $\rho$ region. In particular we note that  the inverse harmonic oscillator potential term, the $-\omega^2 r^2$ term, which drops out in the low energy dynamics, should not cause an alarm, because there are  normalisation factors multiplying the wave function recalling the specific $r$ dependence of the metric components in $r$ coordinate. For exploring the large $r$ far region it is more convenient to use the original $\rho$ coordinate in terms of which the large $\rho$ metric has the standard form of a flat 4d space. Then, the equation becomes a Schr\"{o}dinger type equation for the function $R_{lk\omega}/\rho$ and  the first and leading term in the potential becomes $-\omega^2$ without any $\rho$ dependence; and hence in the large $\rho$ the wave function behaves like $\frac{e^{i\omega\rho}}{\rho}$.} This is analogous to what happens in the usual D$_p$-brane case in the decoupling limit \cite{Oz-Ita-Alishah}. In short, the condition for the near horizon limit to be a decoupling limit is $\omega(k+\sqrt{pq}\omega)>0$.

\subsection{Resolution of  pinching AdS$_3$ orbifold and the EVH/CFT}

Appearance of the \ads{3} throat in the  near horizon of  EVH black holes is very suggestive of existence of a 2d CFT dual to physics on this geometry. One may use standard Brown-Hennueax analysis \cite{Brown-Henneaux} to read the central charge of the proposed 2d CFT from the gravity considerations.
In order to do this one can use the reduction to 3d gravity discussed in section 4.1.3 and that $c_{B.H.}=3R_{AdS3}/(2G_3)=3R^2_{AdS3}/(4G_4)$ for EVH KK black hole. This argument, however, has some caveats: What we obtain in the near horizon is not a round \ads{3}, it is a \emph{pinching orbifold} of \ads{3}. So, we first need to have proposals for ``resolving the pinching orbifold''.

To this end, we adopt and expand on the idea outlined in \cite{EVH-BTZ} (see also \cite{Martinec-McElgin}). We note that in the EVH limit, by definition the area of horizon is vanishing, and this leads to a vanishing entropy for the black hole. As such, one would not expect to have a well-defined gravity picture, e.g. the higher order corrections to the geometry will not be under control.

To remove this problem, we note that Bekenstein-Hawking entropy is the ratio of horizon area to four times Newton constant, and hence if we scale the Newton constant to zero in the same rate as horizon area, in the KK case i.e. $G_4\sim \epsilon\to 0$, we obtain a system with finite entropy. Therefore, we propose to accompany the already ``double scaling near EVH near horizon limit'' of the previous sections by
\be\label{G4-scaling}
G_4=\epsilon \ell^2\,,\qquad \ell,\ R_{AdS3}=fixed\,.
\ee
In this limit the entropy of the near EVH black hole \eqref{near-EVH-entropy} remains finite. The temperature of the black hole is a dimensionful quantity and the dimensionless quantity $T_4\cdot R^3_{AdS3}/G_4$ (\emph{cf.} \eqref{4d-temperature}), which is also proportional to the 3d BTZ temperature, remains finite in this limit. We stress that with the addition of \eqref{G4-scaling} all of our earlier discussions and results of previous sections regarding the near horizon limits and discussions on the entropy and entropy function remains intact.

Scaling $G_4\to 0$ implies vanishing $G_3$ \eqref{G3-G4} and hence the Brown-Henneaux central charge blows up, $c\sim 1/\epsilon$. As we will argue, this is indeed what one needs for ``resolving the pinching orbifold'' \cite{EVH-BTZ}: The pinching \ads{3} orbifold  may be understood as the near horizon limit of massless BTZ in which both of the left and right moving sectors in the dual CFT have decoupled; in  the limit their mass gaps has been sent to infinity as $1/\epsilon$ . On the other hand the mass gap of a (2d) CFT is of order inverse of the central charge $1/c$. Therefore, if we define the near horizon limit such that it also involves a scaling in $c$, the \emph{true} physical mass gap remains finite and the theory will have a non-trivial physical content. For a more detailed discussion we refer the reader to \cite{EVH-BTZ}.

From the 3d (or 4d) gravity viewpoint one can argue that \cite{Martinec-McElgin} the ``physical (Brown-Henneaux) central charge'' for an \ads{3}/$Z_K$ orbifold is indeed
\be
c_K=\frac{3R_{AdS3}}{2G_3}\cdot \frac1K.  \nonumber
\ee
This latter may be understood because the ``effective'' 3d Newton constant for an \ads{3}/$Z_K$ geometry, or any quotient of thereof, is in fact $G_3\cdot K$. \footnote{The factor of $K$ comes from integration over the angular direction of the \ads{3}/$Z_K$. And in our pinching orbifold case $K=1/\epsilon$.} In summary, our proposed
2d CFT, after resolution of the pinching orbifold singularity, has a finite central charge $c$
\be\label{2d-CFT-central-charge}
c=\frac{3R^2_{AdS3}}{2\ell^2}\,.
\ee

Having identified the central charge, the next item in our EVH/CFT dictionary is to relate the scaling dimensions of the proposed 2d CFT, $L_0,\ \bar L_0$ with the quantum numbers and parameters of the near horizon, near EVH geometry, i.e. the BTZ mass and angular momentum or $r_\pm$ \eqref{r-pm}. We recall that, as discussed earlier, once we specified the EVH point, by definition, moving  ``slightly'' away from this point in the parameter space and exciting the EVH black hole geometry, can only happen in two directions and determined by only two parameters, which have been made manifest in the $r_\pm$ of the near EVH near horizon geometry.
%For higher dimensional EVH cases, one of course can envisage more general cases which we will comment upon in the next section.
Besides the thermal excitations of the CFT (the BTZ black holes) we can have excitations corresponding to (quantum) fields turned on and propagate on the EVH background.

Either of these two excitations, the BTZ-type thermal excitations of the EVH geometry and other fields on the EVH background can be used to read off $L_0$ and $\bar L_0$ of the dual CFT.
\begin{itemize}
\item In view of the discussions of previous subsection, let us start with the second one. In the gravity side, and for the ``low energy'' sector which is what is left in the $G_4\to 0$ limit for $l=0$ sector the excitations are only specified by two parameters, $k$ and $\omega$. Let us then focus on the oscillator phase of \eqref{Psi-ansatz} and rewrite it in terms of the near horizon coordinates
\be\label{m-omega-psi-tau}
k\phi+\omega t= \frac{1}{\epsilon}\left[ k\psi+\tau(k+2\sqrt{pq} \omega)\right]\,.
\ee
It is then natural, as we always do, to identify the coefficient of $\psi$ by $J\equiv L_0-\bar L_0$ and the coefficient of $\tau$ by $\Delta\equiv L_0+\bar L_0$. This, however, should be modified in view of the above $G_3\to 0$ scaling. As discussed in \cite{EVH-BTZ}, this should be accompanied by a $1/\epsilon$ scaling. Therefore, we identify
\be\label{L0-L0bar}
L_0= \sqrt{pq}\omega +k\,,\qquad \bar L_0= \sqrt{pq}\omega\,.
\ee
With this identification, the condition for the potential to develop an infinite barrier at $\rho=0$, i.e. $\omega(\sqrt{pq}\omega+k)>0$, implies the positivity of the spectrum of $L_0$ and $\bar L_0$. This latter is nothing but the unitarity condition of the dual 2d CFT. This is what one would have intuitively expected: The near horizon geometry has a unitary description if the wave functions cannot penetrate in or out of the near horizon region. So, the decoupling condition has a very natural (and essential) appearance in the 2d CFT side.
\item The identification of $L_0$ and $\bar L_0$ in terms of the BTZ parameters can be done in the standard way
    \cite{EVH-BTZ}, i.e.
\be\label{L0-barL0}%
L_0=\frac{c}{24}\left(\frac{r_++r_-}{R_{AdS3}}\right)^2\,,\qquad
\bar L_0=\frac{c}{24}\left(\frac{r_+-r_-}{R_{AdS3}}\right)^2\,,
\ee%
where $c$ is the central charge given in \eqref{2d-CFT-central-charge}. The BTZ black hole is then a thermal state
in the 2d CFT specified above at temperature $T_{BTZ}= \frac{r_+^2-r_-^2}{2\pi r_+ }$. With this identification and recalling our earlier discussions, it is then obvious that the Cardy formula which produces the BTZ black hole entropy, recalling \eqref{3d-entropy} and \eqref{4d-entropy}, will also correctly reproduce the near EVH black hole entropy.
\end{itemize}

\section{Discussion and outlook}

In this paper we have considered a specific class of extremal black holes, the Extremal Vanishing Horizon (EVH) black holes. In particular we focused on the 4d EVH black holes. We have two main results: 1) The near horizon limit of any EVH black hole has an \ads{3} throat and, 2) the EVH/CFT correspondence: gravity on near horizon EVH geometry is dual to or described by a 2d CFT. Here we summarize and discuss further these two points and the possible extensions  of these ideas/results.

\paragraph{Near horizon of EVH black holes have \ads{3} throats.}
We have shown that for any 4d EVH black hole the near horizon geometry develops an \ads{3} throat. This \ads{3} is however, generically a \emph{pinching} \ads{3} $\equiv$ \ads{3}$/Z_K,\ Z\to\infty$. Furthermore, we showed that near horizon limit of \emph{near EVH} black holes has a \emph{pinching} BTZ factor. We note that this behaviour seems generic to non-BPS EVH black holes. As suggested by the example of three-charge \ads{4} EVH black holes reviewed in section 4.2, for the BPS EVH black holes we find an \ads{3} throat without the pinching. As the first future direction we outline here one may try to make the above statement about BPS vs. non-BPS EVH black holes more precise.

\paragraph{Further evidence for EVH/CFT.} Appearance of \ads{3} factor in the near horizon geometry is a good indication for trying to establish the EVH/CFT, especially noting the fact that our near horizon limit is indeed a decoupling limit (\emph{cf.} discussions of section 5.1). The main obstacle in the way of EVH/CFT is the pinching \ads{3} issue. To resolve this we proposed to accompany the near EVH near horizon limit by a particular $G_N\to 0$ limit. Explicitly, we proposed the following \emph{triple scaling limit}:
\be\label{triple-scaling}
A_h,\ T, G_4\to 0\,,\qquad A_h/T\ \mathrm{and}\ A_h/G_4\ \mathrm{held\ fixed}\,.
\ee
The above proposal, recalling the standard \ads{3}/CFT$_2$ duality, implies  a particular duality between 2d CFT's on a cylinder and its orbifold:
\begin{center}
2d CFT with central charge $c$ on cylinder $R\times S^1$ is dual to 2d CFT with central charge $cK$ on $R\times S^1/Z_K$ \emph{in the $K\to\infty$ limit}.
\end{center}
For finite $K$ and some particular cases (see \cite{EVH-BTZ} for more details) the above statement is implied by the U-duality on the D1-D5-P system. However, for general $K$, and in particular for large $K$, pinching orbifold, cases we cannot rely on string theory dualities. Here we just make some remarks about the validity of the above statement. Perhaps, preliminary arguments can come from considerations based on Virasoro algebra and some features of BTZ black holes \cite{EVH-BTZ}. Establishing/proving the above statement for 2d CFT's and their pinching orbifolds, is an interesting project on its own, and is of course a key step in establishing our EVH/CFT.

\paragraph{Connection between EVH/CFT and Kerr/CFT.} Recalling that in the Kerr/CFT we are dealing with a chiral CFT, a possible connection between the two can come along the lines of \cite{DLCQ-CFT}: The EVH/CFT in the DLCQ description reproduces Kerr/CFT. For the above to work one should, however, extend the validity of our EVH/CFT proposal beyond the strict near EVH region. In other words, generic Extremal black holes may be viewed as excitations above the EVH black hole, when only e.g. left sector of the dual 2d CFT has been excited. As the first check for this proposal one can show that  for the cases where in the parameter space of the black hole we have an EVH hypersurface, the central charge of 2d CFT in our EVH/CFT and that of Kerr/CFT are equal. We have checked \footnote{This has been done in collaboration with Joan Sim\'on.} this and indeed we get the same central charge for the cases discussed in \cite{Ext/CFT}. \footnote{Of course, in cases where the Kerr/CFT provides the possibility of two chiral CFT's with two different central charges we can only reproduce the one which remains finite in the EVH limit. This point has also been noted in \cite{Terashima}.} Some preliminary steps in making the EVH/CFT vs. Kerr/CFT connection has been taken in \cite{G-S,Compere,Terashima}.

We also remark that our EVH/CFT, at least for the near EVH region, is hence the sought for  non-extremal extensions of Kerr/CFT, e.g. see \cite{Terashima}.

\paragraph{EVH/CFT and Schwarzchild type black holes?} Here we  comment on the possibility of
understanding generic black holes via EVH/CFT. Unlike the Kerr/CFT, EVH/CFT by construct is able to describe general, non-extremal, excitations above the EVH point. Explicitly, as we discussed excitations above the EVH point are labeled by two quantum numbers related to $L_0$ and $\bar L_0$, \eqref{L0-barL0} and \eqref{L0-L0bar}, and both of these can be non-zero. If the EVH/CFT works for very large $L_0$ and $\bar L_0$ then one would have a setup to discuss generic non-extremal black holes, which in the parameter space of black holes are far from the EVH hypersurface.

The question is then how far from the EVH point (hypersurface) the validity of EVH/CFT can be extended. In the strict ``triple decoupling'' limit \eqref{triple-scaling}, we cannot of course probe beyond the \ads{3} throat into the intermediate or asymptotic flat region (see Figure \ref{NH-figure}) by finite energy probes. In other words, in the black hole parameter space  the strict limit
\eqref{triple-scaling} corresponds to restricting oneself to a region very close to the EVH hypersurface, cutting it off from the other parts of the parameter space.  This is very similar what happens in the standard e.g. D3-brane near horizon decoupling limit, leading to AdS$_5$/CFT$_4$ duality.

\paragraph{Embedding EVH black holes in string theory.} For specific examples one may seek embedding into string/M-theory settings. For the KK case, as pointed out this is possible noting the connection to Taub-NUT and 5d uplifts, to Myers-Perry black holes \cite{myersperry}, and from there make use of the D0-D6 brane constructions \cite{Sen:2005wa,KK-black-hole-3}. For the three-charge \ads{4} EVH black holes, we have an M-theory embedding through three intersecting stacks of M5-brane giants. These constructions could then be used to strengthen our EVH/CFT proposal.

\paragraph{EVH black holes in $d>4?$} As we already discussed, EVH black holes are not  limited to 4d geometries. Our definition, black holes with $A_h,\ T\to 0$ with $A_h/T=$fixed, may be extended to any dimension. In particular, in the five dimensions, 5d Kerr with one spin \cite{bardeenhorowitz} and two-charge \ads{5} black holes of $U(1)^3$ gauged SUGRA \cite{Balasubramanian:2007bs,Fareghbal:2008ar} already fall within our EVH black hole definition \cite{In-progress}. The uplift of KK black hole to 5d discussed in \eqref{5d-uplift} is another example. Our analysis shows that \cite{Joan-unpublished} BPS or non-BPS EVH black holes of \cite{5d-blackholes-sugra}, exhibit the same behaviour as 4d cases discussed above.

\paragraph{EVH black holes vs. EVH black rings.} It is now established  that in higher dimensional geometries we can have (asymptotically flat) black objects with non-spherical horizon topology, e.g. see \cite{Ring-review} and references therein. On the other hand, as we discussed in sections 3 and 4, and depicted in Figure \ref{horizon}, the horizon of  the EVH \emph{black hole}, although still topologically spherical, is geometrically  extremely deformed and in particular in the near horizon limit it essentially becomes a cylinder.  Next, we recall that in the class of black ring solutions
there are ``EVH black rings'', black rings with a single spin.\footnote{We remark that there could be many EVH multi-black ring/saturn solutions, as well as many different EVH black holes, becoming essentially indistinguishable in the near horizon limit. Examples of these cases may be found in \cite{multi-ring}.} One may then wonder e.g. by identifying the end points of this cylinder one may get a (near EVH) black ring and if the horizon topology change\footnote{The topology-changing transitions can also happen in non-extremal black holes/rings. In those cases the transition is presumably not controlled  by  Kerr/CFT type duality. Nonetheless, the EVH/CFT may be in a better position to do so, if the black holes in question are not in far EVH region.}
can happen through the EVH point.

For concreteness let us consider 5d asymptotically flat black solutions to vacuum Einstein equations. In this class extremal black holes and rings are generically specified by two spins. In both black hole and ring branches,  EVH geometries are those with only one spin \cite{bardeenhorowitz,Emparan-Reall}. At this point the space of solutions becomes degenerate and by turning on the other spin one may move to either of the branches. Our discussion here suggests that there is a window (for small deviations around EVH point) that we have the \ads{3} throat and hence a 2d CFT description. The natural question is then how in the 2d CFT description we can distinguish moving to black hole or black ring branches?

To answer this question let us recall the discussion of \cite{DLCQ-CFT,EVH-BTZ,vijay-simon}:

In the 2d CFT we can choose two different vacuum states, 1) the
standard vacuum state with $L_0=\bar L_0=c/24$ with both left and right
sectors having the same mass gap and, 2) the vacuum state of a DLCQ CFT.
In this case although still $L_0=\bar L_0=c/24$, due to the DLCQ, the
mass gap in one sector remains finite while the other one is taken
to infinity. Geometrically, the first one corresponds to usual
massless BTZ and the second to a null self-dual orbifold.

If we approach the EVH point from the black hole side we are in
fact in a situation like vacuum 1) of the CFT, while if we approach
the EVH point from the ring side we find the vacuum 2).
This is compatible with discussions in \cite{G-S,Compere,Terashima}, while sheds new light on the appearance of (null) self-dual \ads{3} orbifold in these settings.

\paragraph{Other extensions to EVH black holes.} In the definition of EVH black holes we chose to keep $A_h/T$ finite.
As we showed this choice, through appearance of the \ads{3} throat, and the EVH/CFT correspondence, is ultimately related to the 2d CFT and the Cardy formula. One can then imagine extensions of the above by defining EVH$_n$ black holes, for which $A_h/T^n$ is kept finite; the $n=1$  being the EVH case we studied here. For $n>1$, again based on the generic property of $n+1$ dimensional CFT's for which $S\propto c\ T^n$ with $c$ being the central charge, one would expect to obtain an \ads{n+2} throat in the near horizon geometry. So far we do not have any example of $n>1$ cases, but it would be interesting to search for.\footnote{We thank Roberto Emparan for correspondence on this point.}

\section*{Acknowledgement}

 M.M.Sh-J. would like to thank Jan de Boer and especially Joan Sim\'on for long discussions and collaborations on very related subjects over the past three years. We would like to thank Roberto Emparan for discussions and comments on the draft.

\appendix

\section{Heun's Equations}
A natural extension of the Riemann p-differential equation is called general Heun's equation and given by
\be
\label{GHE}
H'' + \left( \frac{\gamma}{z}+ \frac{\delta}{z-1} + \frac{\zeta}{z-a}\right) H' +  \frac{\alpha \beta z-\nu}{z (z-1)(z-a)} H =0,
\ee
which has four regular singular points located at $z= 0,1,a, \infty$. Among parameters in above equation, ${\mathrm q}$ falls outside the domain of the usual Riemann classification scheme and it is referred to an auxiliary parameter and the five other satisfy the Fuchsian condition $1+\alpha+ \beta=\gamma +\delta + \zeta$. For our problem, the Heun equation parameters and the parameters of our ansatz solution \eqref{Psi-ansatz} and the background $p,q$ are related as
\bea
&&\gamma =k+1, \;\;\; \delta= k+1,\;\;\;\; \nu= k(k+1)-\frac{2ql(l+1)}{p+q}-\omega^2\frac{q(2p+3q)(p-q)}{p+q},\;\;\; \cr\cr
&&\alpha \beta = \frac{\omega^2 pq(p-q)}{p+q}, \;\;\;    z=\frac{1}{2}(x+1)
\eea

Using the method of Frobenius, one can derive local power series solutions to this equation in the neighbourhood of singular points. These local solutions are normally valid only in a circle which excludes the nearest other singularity, and can be written as a power series. Such solutions are called Frobenius solutions. A local Frobenius solution about a singularity $s_1$ can be continued analytically to a neighbourhood of an adjacent singularity $s_2$ but will not generally coincide with the local solution near $s_2$. The most important solutions to the physical applications are those that are simultaneously local Frobenius solutions about two adjacent singular points. These are referred to as \emph{Heun functions}.

Confluent forms of Heun's differential equation arise when two or more of the regular singularities merge to form an irregular singularity. There are four standard forms, corresponding to four confluence processes

\begin{itemize}
\item {\bf Confluent Heun's equation}
   \end{itemize}

Starting from the general Heun's Equations one can obtain Confluent Heun's equation using the following limit
\be
\beta \rightarrow \beta a, \;\;\;\; \epsilon \rightarrow \epsilon a,  \;\;\;\; \nu \rightarrow \nu a,  \;\;\;\; a \rightarrow \infty
\ee
Therefore the equation {\ref{GHE}} will change to the following equation
\be
\label{CHEQ}
H'' + \left( \frac{\gamma}{z}+ \frac{\delta}{z-1} - {\epsilon}\right) H' +  \frac{\nu-\alpha \beta z}{z (z-1)} H =0,
\ee
which is called Confluent Heun's equation. This has regular singularities at $z=0$ and $1$, and an irregular singularity of rank one at $z=\infty$. Mathieu's functions, spheroidal wave functions, and Coulomb spheroidal functions are special cases of solutions of the confluent Heun equation.

\begin{itemize}
\item {\bf Doubly-Confluent Heun's Equation}
   \end{itemize}
Similarly we can get the other forms of Heun's Equation by merging singularities at $0$ and $\infty$ we obtain
\be
H'' + \left( \frac{\delta}{z^2}+ \frac{\gamma}{z} + 1 \right) H' +  \frac{\alpha z-\nu }{z^2} H =0\,,
\ee
which has irregular singularities at $z=0$  and $z=\infty$, both of rank one.

\begin{itemize}
\item {\bf Biconfluent Heun Equation}
   \end{itemize}
The next one which is called Biconfluent Heun's Equation is given by merging three singularities at $\infty$ we get
\be
H'' + \left( \frac{\delta}{z}+ \delta + z \right) H' +  \frac{\alpha z-\nu }{z} H =0\,.
\ee
This has a regular singularity at $z=0$, and an irregular singularity of rank two at $z=\infty$.

\begin{itemize}
\item {\bf Triconfluent Heun Equation}
\end{itemize}
The last form of Heun's Equation which is given by unifying all singularities at $\infty$ and can be written as
\be
H'' + \left( {\delta}{z} + z^2 \right) H' +  (\alpha z-\nu ) H =0\,.
\ee
This has one singularity, an irregular singularity of rank 3 at $z=\infty$.

\end{document}